\documentclass[twocolumn,showpacs,aps,prl,superscriptaddress,preprintnumbers]{revtex4}

\long\def\inst#1{\par\nobreak\kern 4pt\nobreak
    {\it #1}\par\vskip 10pt plus 3pt minus 3pt}

\usepackage[dvips]{graphicx}
\usepackage{epsf,epsfig,rotating,color,colordvi,relsize,xspace}

\RequirePackage{xspace}

\newcommand{\SlacPub}     {SLAC-PUB-15814}

\newcommand{\BabarPub}     {\babar-PUB-13/016}

\input plb_symbols
\def\pipi   {\ensuremath{\pi\pi}}

\def\lhratio  {\rm BDT}

\def\calB      {\ensuremath{{\cal B}}\xspace}

\def\ellpp       {\ensuremath{\ell'^{+}}\xspace}

\def\Bmeson  {\B-meson}

\def\Dmeson  {\ensuremath{D} meson}

\def\Kmm    {\ensuremath{\Km\mup\mup}}
\def\BtoKmm {\ensuremath{\Bp\to\Kmm}}
\def\Pimm    {\ensuremath{\pim\mup\mup}}
\def\BtoPimm {\ensuremath{\Bp\to\Pimm}}

\def\Rhoee   {\ensuremath{\rhom\ep\ep}}
\def\Rhoemu  {\ensuremath{\rhom\ep\mup}}
\def\Rhomumu {\ensuremath{\rhom\mup\mup}}

\def\Dcee    {\ensuremath{\Dm\ep\ep}}
\def\Dcemu   {\ensuremath{\Dm\ep\mup}}
\def\Dcmumu {\ensuremath{\Dm\mup\mup}}

\def\Piemu {\ensuremath{\pim\ep\mup}}

\def\Kemu  {\ensuremath{\Km\ep\mup}}

\def\BtoRhoee   {\ensuremath{\Bp\to\Rhoee}}
\def\BtoRhoemu  {\ensuremath{\Bp\to\Rhoemu}}
\def\BtoRhomumu {\ensuremath{\Bp\to\Rhomumu}}
\def\BtoDcee    {\ensuremath{\Bp\to\Dcee}}
\def\BtoDcemu   {\ensuremath{\Bp\to\Dcemu}}
\def\BtoDcmumu  {\ensuremath{\Bp\to\Dcmumu}}

\def\BtoKemu  {\ensuremath{\Bp\to\Kemu}}
\def\BtoPiemu {\ensuremath{\Bp\to\Piemu}}

\newcommand{\nbb}        {\mbox{$471\pm3$}}
\newcommand{\KstarKpeeEff}     {\mbox{$11.5\pm0.1$}}
\newcommand{\KstarKpemuEff}    {\mbox{$7.9\pm0.1$}}
\newcommand{\KstarKpmumuEff}   {\mbox{$6.1\pm0.1$}}
\newcommand{\KstarKzeeEff}     {\mbox{$12.3\pm0.1$}}
\newcommand{\KstarKzemuEff}    {\mbox{$8.5\pm0.1$}}
\newcommand{\KstarKzmumuEff}   {\mbox{$5.8\pm0.1$}}
\newcommand{\RhoeeEff}    {\mbox{$12.1\pm0.1$}}
\newcommand{\RhoemuEff}   {\mbox{$10.3\pm0.1$}}
\newcommand{\RhomumuEff}  {\mbox{$7.3\pm0.1$}}
\newcommand{\DceeEff}     {\mbox{$10.2\pm0.1$}}
\newcommand{\DcemuEff}    {\mbox{$7.7\pm0.1$}}
\newcommand{\DcmumuEff}   {\mbox{$5.7\pm0.1$}}

\newcommand{\KemuEff}     {\mbox{$15.2\pm0.1$}}
\newcommand{\PiemuEff}    {\mbox{$16.4\pm0.2$}}

\newcommand{\KstarKpeeNtot}   {\mbox{$63$}}
\newcommand{\KstarKpemuNtot}  {\mbox{$117$}}
\newcommand{\KstarKpmumuNtot} {\mbox{$85$}}
\newcommand{\KstarKzeeNtot}   {\mbox{$91$}}
\newcommand{\KstarKzemuNtot}  {\mbox{$172$}}
\newcommand{\KstarKzmumuNtot} {\mbox{$98$}}
\newcommand{\RhoeeNtot}      {\mbox{$411$}}
\newcommand{\RhoemuNtot}      {\mbox{$1651$}}
\newcommand{\RhomumuNtot}     {\mbox{$936$}}
\newcommand{\DceeNtot}        {\mbox{$401$}}
\newcommand{\DcemuNtot}       {\mbox{$549$}}
\newcommand{\DcmumuNtot}      {\mbox{$229$}}
\newcommand{\KemuNtot}        {\mbox{$117$}}
\newcommand{\PiemuNtot}       {\mbox{$464$}}

\newcommand{\KstarKpeeNsig}      {\mbox{$3.8\pm3.3$}}
\newcommand{\KstarKpemuNsig}     {\mbox{$-1.9\pm4.7$}}
\newcommand{\KstarKpmumuNsig}    {\mbox{$2.3\pm1.8$}}
\newcommand{\KstarKzeeNsig}      {\mbox{$0.8\pm3.9$}}
\newcommand{\KstarKzemuNsig}     {\mbox{$-5.1\pm2.6$}}
\newcommand{\KstarKzmumuNsig}    {\mbox{$2.0\pm1.8$}}
\newcommand{\RhoeeNsig}       {\mbox{$-2.1\pm5.7$}}
\newcommand{\RhoemuNsig}      {\mbox{$4.6\pm11.4$}}
\newcommand{\RhomumuNsig}     {\mbox{$2.9\pm6.8$}}
\newcommand{\DceeNsig}        {\mbox{$3.9\pm4.8$}}
\newcommand{\DcemuNsig}       {\mbox{$1.1\pm3.2$}}
\newcommand{\DcmumuNsig}      {\mbox{$-1.7\pm2.5$}}
\newcommand{\KemuNsig}        {\mbox{$5.5\pm3.5$}}
\newcommand{\PiemuNsig}       {\mbox{$3.8\pm3.5$}}

\newcommand{\KstarKpeeBFsyst}   {\mbox{$0.2$}}
\newcommand{\KstarKpemuBFsyst}  {\mbox{$0.4$}}
\newcommand{\KstarKpmumuBFsyst} {\mbox{$0.2$}}
\newcommand{\KstarKzeeBFsyst}   {\mbox{$0.2$}}
\newcommand{\KstarKzemuBFsyst}  {\mbox{$0.7$}}
\newcommand{\KstarKzmumuBFsyst} {\mbox{$0.9$}}
\newcommand{\RhoeeBFsyst}      {\mbox{$0.1$}}
\newcommand{\RhoemuBFsyst}     {\mbox{$0.2$}}
\newcommand{\RhomumuBFsyst}    {\mbox{$0.3$}}
\newcommand{\DceeBFsyst}       {\mbox{$1.5$}}
\newcommand{\DcemuBFsyst}      {\mbox{$1.1$}}
\newcommand{\DcmumuBFsyst}     {\mbox{$0.9$}}
\newcommand{\KemuBFsyst}     {\mbox{$0.1$}}
\newcommand{\PiemuBFsyst}    {\mbox{$0.1$}}

\newcommand{\KstarKpeeSigfull}     {\mbox{$1.2$}}
\newcommand{\KstarKpemuSigfull}    {\mbox{$0.0$}}
\newcommand{\KstarKpmumuSigfull}   {\mbox{$1.3$}}
\newcommand{\KstarKzeeSigfull}     {\mbox{$0.3$}}
\newcommand{\KstarKzemuSigfull}    {\mbox{$0.0$}}
\newcommand{\KstarKzmumuSigfull}   {\mbox{$1.0$}}
\newcommand{\RhoeeSigfull}      {\mbox{$0.0$}}
\newcommand{\RhoemuSigfull}     {\mbox{$0.4$}}
\newcommand{\RhomumuSigfull}    {\mbox{$0.5$}}
\newcommand{\DceeSigfull}    {\mbox{$1.0$}}
\newcommand{\DcemuSigfull}   {\mbox{$0.5$}}
\newcommand{\DcmumuSigfull}  {\mbox{$0.0$}}
\newcommand{\KemuSigfull}    {\mbox{$1.8$}}
\newcommand{\PiemuSigfull}   {\mbox{$1.2$}}

\newcommand{\KstarKpeeBF}   {\mbox{$2.1 \pm1.8\pm \KstarKpeeBFsyst$}}
\newcommand{\KstarKpemuBF}  {\mbox{$-1.5\pm3.8\pm \KstarKpemuBFsyst$}}
\newcommand{\KstarKpmumuBF} {\mbox{$2.0 \pm1.8\pm \KstarKpmumuBFsyst$}}
\newcommand{\KstarKzeeBF}   {\mbox{$0.6 \pm2.9\pm \KstarKzeeBFsyst$}}
\newcommand{\KstarKzemuBF}  {\mbox{$-6.0\pm2.8\pm \KstarKzemuBFsyst$}}
\newcommand{\KstarKzmumuBF} {\mbox{$3.1 \pm2.9\pm \KstarKzmumuBFsyst$}}

\newcommand{\RhoeeBF}       {\mbox{$-0.4\pm1.0\pm \RhoeeBFsyst$}}
\newcommand{\RhoemuBF}      {\mbox{$1.0 \pm2.4\pm \RhoemuBFsyst$}}
\newcommand{\RhomumuBF}     {\mbox{$0.9 \pm2.0\pm \RhomumuBFsyst$}}
\newcommand{\DceeBF}        {\mbox{$8.8 \pm8.6\pm \DceeBFsyst$}}
\newcommand{\DcemuBF}       {\mbox{$3.4 \pm9.4\pm \DcemuBFsyst$}}
\newcommand{\DcmumuBF}      {\mbox{$-6.5\pm9.9\pm \DcmumuBFsyst$}}
\newcommand{\KemuBF}        {\mbox{$0.6 \pm0.5\pm \KemuBFsyst$}}
\newcommand{\PiemuBF}       {\mbox{$0.5 \pm0.5\pm \PiemuBFsyst$}}

\newcommand{\KstarKpeeUL}     {\mbox{$5.1$}}
\newcommand{\KstarKpemuUL}    {\mbox{$6.5$}}
\newcommand{\KstarKpmumuUL}   {\mbox{$7.0$}}

\newcommand{\KstarKzeeUL}     {\mbox{$6.0$}}
\newcommand{\KstarKzemuUL}    {\mbox{$4.2$}}
\newcommand{\KstarKzmumuUL}   {\mbox{$9.8$}}

\newcommand{\RhoeeUL}    {\mbox{$1.7$}}
\newcommand{\RhoemuUL}   {\mbox{$4.7$}}
\newcommand{\RhomumuUL}  {\mbox{$4.2$}}
\newcommand{\DceeUL}     {\mbox{$26$}}
\newcommand{\DcemuUL}    {\mbox{$21$}}
\newcommand{\DcmumuUL}   {\mbox{$17$}}
\newcommand{\KemuUL}     {\mbox{$1.6$}}
\newcommand{\PiemuUL}    {\mbox{$1.5$}}

\begin{document}
\title{
{\rm
  \begin{flushleft}
%      \mbox{\normalsize \BabarPub\ \hspace{3cm}  \ArXiv\ \BaBarType\ \BaBarNumber\ Ver. \BaBarVersion}
     {\normalsize \BabarPub}  \\ {\normalsize \SlacPub}
  \end{flushleft}
}
 \large \bfseries \boldmath Search for lepton-number violating $\Bp \to X^- \ellp\ellpp$ decays
}

%% author list as of 06-Sep-2013 (335 authors)
%
\author{J.~P.~Lees}
\author{V.~Poireau}
\author{V.~Tisserand}
\affiliation{Laboratoire d'Annecy-le-Vieux de Physique des Particules (LAPP), Universit\'e de Savoie, CNRS/IN2P3,  F-74941 Annecy-Le-Vieux, France}
\author{E.~Grauges}
\affiliation{Universitat de Barcelona, Facultat de Fisica, Departament ECM, E-08028 Barcelona, Spain }
\author{A.~Palano$^{ab}$ }
\affiliation{INFN Sezione di Bari$^{a}$; Dipartimento di Fisica, Universit\`a di Bari$^{b}$, I-70126 Bari, Italy }
\author{G.~Eigen}
\author{B.~Stugu}
\affiliation{University of Bergen, Institute of Physics, N-5007 Bergen, Norway }
\author{D.~N.~Brown}
\author{L.~T.~Kerth}
\author{Yu.~G.~Kolomensky}
\author{M.~J.~Lee}
\author{G.~Lynch}
\affiliation{Lawrence Berkeley National Laboratory and University of California, Berkeley, California 94720, USA }
% ...
\collaboration{The \babar\ Collaboration}
\noaffiliation

% comment out before submission
%\date{\today}

\begin{abstract}
We report on a search for eleven lepton-number violating processes
$\Bp\to X^- \ellp\ellpp$ with $X^- = \ensuremath{K^{-}}$, $\pi^-$,
$\rho^-$, $\ensuremath{K^{*-}}$, or \Dm\ and $\ellp/\ellpp=e^+$ or
$\mup$, using a sample of \nbb\ million \BB\ events collected with the
\babar\ detector at the \pep2\ $\epem$ collider at the SLAC National
Accelerator Laboratory. We find no evidence for any of these modes and
place 90\% confidence level upper limits on their branching fractions
in the range $(1.5-26)\times 10^{-7}$.
\end{abstract}


\maketitle

%---------------------------------------------------------------------
In the Standard Model (SM), lepton-number conservation holds in
low-energy collisions and decays but it can be violated in high-energy
or high-density interactions~\cite{bib:sphaleron}. The observation of neutrino
oscillations~\cite{bib:neutrinos} indicates that neutrinos have mass
and, if the neutrinos are of the Majorana type, the neutrino and
antineutrino are the same particle and processes that involve
lepton-number violation become possible~\cite{bib:majorana}. Many
models beyond the SM predict that lepton-number is
violated, possibly at rates approaching those accessible with current
data~\cite{bib:atre}. Lepton-number violation is also a
necessary condition for leptogenesis as an explanation of the baryon
asymmetry of the universe~\cite{bib:leptogenesis}.

Following recent results from \lhcb~\cite{bib:lnv_lhcb},
\babar~\cite{bib:lnv_babar} and \belle~\cite{bib:lnv_belle}, there has
been interest in the possibility of measuring the lepton-number
violating (LNV) processes $B^+\rightarrow X^- \ellp\ellpp$, where
$X^-$ is a charged hadronic particle or resonance, and
$\ellp/\ellpp=\ep$ or $\mup$~\cite{bib:charge}. Earlier searches for
these decays by the CLEO collaboration have produced 90\% confidence
level (C.L.)  upper limits on the branching fractions in the range
$(1.0-8.3) \times 10^{-6}$~\cite{bib:cleo}. The LHCb collaboration
reported 95\% C.L. upper limits on the branching fractions
$\calB(\BtoKmm) < 5.4 \times 10^{-8}$ and $\calB(\BtoPimm) < 1.3
\times 10^{-8}$~\cite{bib:lnv_lhcb}. The Belle collaboration places
90\% C.L. upper limits on the branching fractions
$\calB(\Bp\to\Dm\ellp\ellpp)$ in the range $(1.1-2.6)\times
10^{-6}$~\cite{bib:lnv_belle}.

We report here on a search for $\Bp\to X^- \ellp\ellpp$ with $X^- =
\ensuremath{K^{-}}$, $\pi^-$, $\rho^-$, $\ensuremath{K^{*-}}$, or \Dm\
and $\ellp/\ellpp=e^+$ or $\mup$. We exclude the four combinations
previously measured by the \babar\
collaboration~\cite{bib:lnv_babar}. We use a data sample of \nbb\
million \BB\ pairs, equivalent to an integrated luminosity of
429\invfb~\cite{bib:lumi}, collected at the $\FourS$ resonance with
the \babar\ detector at the \pep2\ asymmetric-energy $\epem$ collider
at the SLAC National Accelerator Laboratory. The \epem\ center-of-mass
(CM) energy is $\sqrt{s} = 10.58$\gev, corresponding to the mass of
the \FourS\-resonance (on-resonance data). The \babar\ detector is
described in detail in Ref.~\cite{BaBarDetector}.

Monte Carlo (MC) simulation is used to identify the background
contamination, calculate selection efficiencies, evaluate systematic
uncertainties, and to cross-check the selection procedure.  The signal
channels are simulated by the EvtGen~\cite{bib:evtgen} package using a
three-body phase space model. We also generate light quark \qqbar\
continuum events ($\epem\to\qqbar$, $q=u,d,s$), charm $\epem\to\ccbar$
continuum events, $\epem\to\mup\mun(\gamma)$, Bhabha
elastic \epem\ scattering, \BB\ background, and two-photon
events~\cite{bib:twophoton}.  Final-state radiation is provided by
\photos~\cite{bib:photos}. The detector response is simulated with
\geantfour~\cite{bib:geant4}, and all simulated events are
reconstructed in the same manner as data.

Particle identification is applied to all charged tracks.  The charged
pions and kaons are identified by measurements of their energy loss in
the tracking detectors, and the number of photons and the Cherenkov angle
recorded by the ring-imaging Cherenkov detector. These measurements
are combined with information from the electromagnetic calorimeter and
the muon detector to identify electrons and
muons~\cite{BaBarDetector}.

We select events that have four or more charged tracks, at least two
of which must be identified as leptons. The ratio of the
second-to-zeroth Fox-Wolfram moments~\cite{bib:fox} of the event must
be less than 0.5 and the two charged leptons must have the same sign
and momenta greater than $0.3$\gevc\ in the laboratory frame. The
separation along the beam axis between the two leptons at their closest
approach to the beamline is required to be less than 0.2\cm. The combined
momentum of the \ensuremath{\ellp\ell'^{+}}\xspace pair in the CM
system must be less than $2.5$\gevc. Electrons and positrons from photon
conversions are removed, where photon conversion is indicated by
electron-positron pairs with an invariant mass less than 0.03\gevcc
and a production vertex more than 2\cm\ from the beam axis.

The \Kstarm\ is reconstructed through its decay to $\KS\pim$ and
$K^-\pi^0$; the \rhom\ and \Dm are reconstructed through their decays
to $\pi^-\pi^0$ and $K^+\pi^-\pi^-$, respectively. The photons from 
the \piz\ must have an energy greater than 0.03\gev, and the \piz is
required to have an energy greater than 0.2\gev, both measured in the
laboratory frame. The reconstructed \piz\ invariant mass must be
between 0.12 and 0.16\gevcc. The invariant mass of the \rhom\ is
required to be between 0.470 and 1.07\gevcc. The \KS\ must have an
opening angle $\theta$ between its flight direction (defined as the
vector between the \B\ meson and \KS\ vertices) and its momentum vector
such that $\cos\theta> 0.999$, a transverse flight distance greater
than 0.2\cm, a lifetime significance $\tau/\sigma_{\tau}>10$, and a
reconstructed invariant mass between 0.488 and 0.508\gevcc. We
require the \Dm\ invariant mass to be between 1.835 and
1.895\gevcc. The invariant mass ranges are chosen to be wide enough to
allow the background event distributions to be modeled.

The two leptons are combined with either a resonance candidate or a
charged track to form a \Bmeson\ candidate. For muon modes, the
invariant mass of each combination of a muon and a charged
track from the \Bmeson\ candidate must be outside the region
$3.05<m_{\ellp h^-}<3.13$\gevcc.  This rejects events where a muon
from a $\jpsi$ decay is misidentified as a pion. The probability to
misidentify a pion as a muon is approximately 2\% and to misidentify
as an electron is less than $0.1\%$.

We measure the kinematic variables $\mes=\sqrt{(s/2 + {\bf p}_0 \cdot
{\bf p}_B)^2/E^2_0 - {\bf p}^2_B}$ and $\Delta E = E_B^* -
\sqrt{s}/2$, where $(E_0, {\bf p}_0)$ is the four-momentum of the CM system
and ${\bf p}_B$ is the \B candidate momentum vector, both measured in
the laboratory frame, $\sqrt{s}$ is the total CM energy, and $E_B^*$
is the energy of the \B candidate in the CM system.  For signal
events, the \mes\ distribution peaks at the \B meson mass with a
resolution of about 2.5\mevcc, and the \DeltaE\ distribution peaks
near zero with a resolution of about 20\mev. The \B candidate is
required to be in the kinematic region $5.240 <\mes < 5.289 \gevcc$.
The \DeltaE\ range depends on the mode but always satisfies $|\DeltaE|<0.3$\gev.

The backgrounds arising from \qqbar, \ccbar, and \BB\ events are
suppressed through the use of a boosted decision tree discriminant
(BDT)~\cite{bib:bdt}. The BDT has nine inputs: the ratio of the
second- and zeroth-order Fox-Wolfram moments based on the magnitude of
the momentum of all neutral clusters and charged tracks in the
rest-of-the-event (ROE) not associated with the \B\ candidate; the
absolute value of the cosine of the angle between the \B\ momentum and
the beam axis in the CM frame; the absolute value of the cosine of the
angle between the \B\ thrust axis~\cite{bib:thrust} and the beam axis
in the CM frame; the absolute value of the cosine of the angle between
the \B\ thrust axis and the thrust axis of the ROE in the CM frame;
the output of the flavor tagging algorithm~\cite{bib:tag}; the
separation along the beam axis between the two leptons at their points
of closest approach to the beamline; the missing energy in the CM; the
momentum of the lepton pair in the CM; and the boost-corrected
proper-time difference between the decays of the two \B\ mesons
divided by its variance. The second \B\ meson is formed by creating a
vertex from the remaining tracks that are consistent with originating
from the interaction region.  The discriminant is trained for each
signal mode using on-resonance data with $\mes<5.27\gevcc$ together
with samples of simulated signal and background events. We compare the
distributions of the data and the simulated background variables used
as input to the BDTs and confirm that they are consistent.

\begin{figure}[ht!] 
\begin{center} 
\begin{tabular}{c}
    \epsfig{file=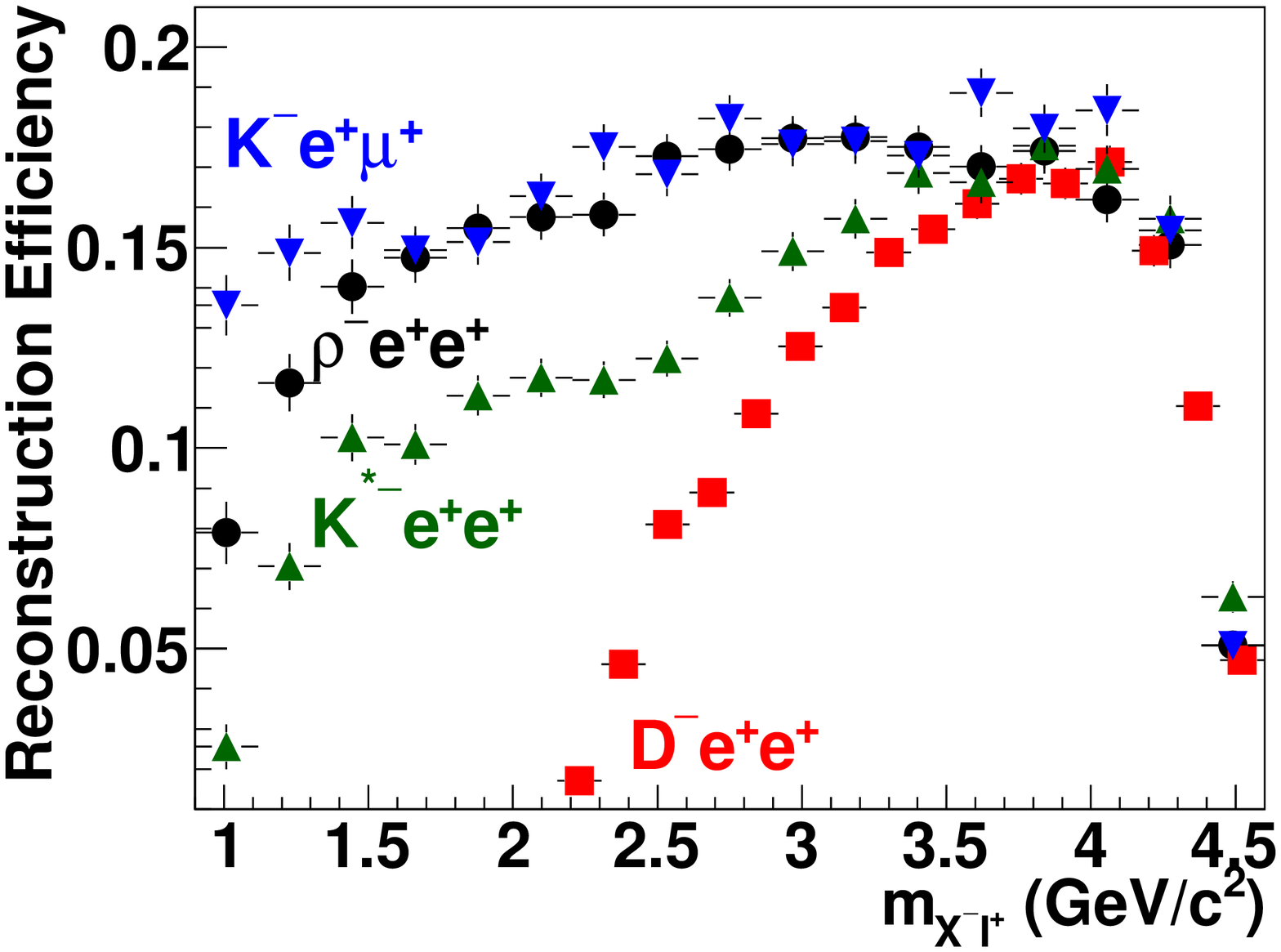,width=0.90\columnwidth}
\end{tabular}
\caption{Reconstruction efficiency as a function of $m_{X^-\ellp}$ for
  four of the modes: \BtoKemu\ (down triangle), \BtoRhoee\ (circle),
  $\Bp\to\Kstarm(\to\KS\pim)\ep\ep$ (up triangle), and
  \BtoDcee\ (square).}
\label{fig:effic}
\end{center} 
\end{figure}

After the application of all selection criteria, some events will
contain more than one reconstructed \B candidate. From the simulation
sample, we estimate that the fraction of signal events with a \piz\
that have more than one candidate is 30\%, 13\% for signal events with
a \KS, and less than 6\% for signal events where the lepton pair is
combined with a \Dm, \Km, or \pim. We select the most
probable \B candidate from among all the candidates in the event using
the $\chi^2$ from the \B\ candidate vertex fit.  Averaged over all
simulated signal events, the correct \B candidate is selected with an
accuracy of more than 83\% for the signal events with a \piz, greater
than 96\% for signal events with a \KS, and over 99\% for signal
events where the lepton pair is combined with a
\Dm, \Km, or \pim.  The final event selection efficiency for simulated
signal is between 6\% and 16\%, depending on the final state.
Figure~\ref{fig:effic} shows the variation of the selection efficiency
as a function of the reconstructed invariant mass $m_{X^-\ellp}$,
calculated for both leptons, for
four of the modes. The remaining modes have similar distributions.

For each mode, we extract the signal and background 
yields from the data with an unbinned maximum likelihood (ML) fit using

\begin{equation}
  \label{eq:ml}
{\mathcal L} = \frac{1}{N!}\exp{\left(-\sum_{j}n_{j}\right)}
\prod_{i=1}^N\left[\sum_{j}n_{j}{\mathcal
    P}_{j}(\vec{x}_i;\vec{\alpha}_j)\right]\!,
\end{equation}

\noindent where the likelihood for each event candidate $i$ is the sum
of $n_j {\cal P}_j(\vec x_i; \vec \alpha_j)$ over two categories $j$:
the signal mode $\Bp\to X^-\ellp\ellpp$ (including the small number of
misreconstructed signal candidates) and background, as will be discussed.
For each category $j$, ${\cal P}_j(\vec x_i; \vec \alpha_j)$ is the
product of the probability density functions (PDFs) evaluated for the
$i$-th event's measured variables $\vec x_i$.  The number of events for
category $j$ is denoted by $n_j$ and $N$ is the total number of events in
the sample. The quantities $\vec \alpha_j$ represent the parameters
describing the expected distributions of the measured variables for
each category~$j$. Each discriminating variable $\vec x_i$ in the
likelihood function is modeled with a PDF, where the parameters $\vec
\alpha_j$ are extracted from MC simulation or
on-resonance data with \mes $< 5.27$\gevcc. The variables $\vec x_i$
used in the fit are \mes, \DeltaE, and the multivariate discriminant
\lhratio\ output; for modes involving a resonance, the resonance invariant
mass is included as a fourth variable. The linear correlations
between the four variables are found to be typically 4\%-9\% for
simulated signal modes. Only $\Bp\to\Dm\ep\ep$ shows a larger
correlation, between the invariant mass and \DeltaE, due to the
occasional Bremstrahlung energy loss from the electrons. 
We take each ${\cal P}_j$ to be the product of the PDFs for the
separate variables and treat any correlations in the variables later
as a source of systematic uncertainty.

MC simulations show that the \qqbar, \ccbar, and \BB\ backgrounds have
similar distributions in the four variables after the selection
criteria have been applied and we therefore use a single background
parameterization. An ARGUS parameterization~\cite{ArgusShape} is used
to describe the \mes\ distribution. For \DeltaE, a first- or
second-order polynomial is used or, for modes with a \Dmeson, a
Cruijff function~\cite{bib:cruijff}. The multivariate discriminant
\lhratio\ output is fitted using a non-parametric kernel estimation
KEYS algorithm~\cite{bib:keys}. The mass distributions for modes with
a resonance are fitted with a first-order polynomial, together with a
Gaussian function if the resonance is present in the backgrounds.
 
For the signal, the \mes\ distribution is parameterized with a Crystal
Ball function~\cite{bib:cryball}. A Crystal Ball function is also used
for \DeltaE, together with a first-order polynomial for modes with a
\piz. The multivariate discriminant \lhratio\ output is taken directly
from the MC distribution using a histogram. For the resonances, the
signal masses are parameterized with two Gaussians, a relativistic
Breit-Wigner function, and a Gounaris-Sakurai function~\cite{bib:gs}
for the \Dm, \Kstarm, and \rhom, respectively. The free parameters in
the ML fit are the signal and background event yields, the slope of
the background \mes\ distribution, and the polynomial parameters of
the background \DeltaE\ and mass distributions.

We test the performance of the fits to $\Bp\to X^-\ellp\ellpp$ by
generating ensembles of MC datasets from both the PDF distributions
and the fully simulated MC events; in the latter case, the
correlations between the variables are correctly simulated. We
generate and fit 10,000 datasets with the numbers of signal and
background events allowed to fluctuate according to a Poisson
distribution. The signal yield bias in the ensemble of fits is between
$-0.3$ and 1.2 events, depending on the mode, and this is subtracted
from the yield obtained from the data.

As a cross-check of the background PDFs, we perform a fit to a
simulated background sample, with the same number of events as the
on-resonance data sample. The number of fitted signal events is
compatible with zero for all modes. We also perform a blinded fit to
the on-resonance data for each mode and confirm that the distributions
of the background events are reproduced by the background PDFs. Events
are identified as background if ${\cal P}_{\rm
bck}/({\cal P}_{\rm bck}+{\cal P}_{\rm sig})>0.9$, where ${\cal
P}_{\rm sig}$ and ${\cal P}_{\rm bck}$ are computed for each event for
the signal and background, respectively, without the use of the
variable under consideration.

% earlier babar results
\babar has previously published, using a different selection
technique, four measurements of LNV in $\Bp\to h^-\ellp\ellp$ where
$h^-=\Km$ or \pim, and \ellp\ellp = $e^+e^+$ or
$\mu^+\mu^+$~\cite{bib:lnv_babar}. To validate the analysis reported
here and as a cross-check only, we repeat the previous measurements,
using the selection criteria described in this article. The
reconstruction efficiencies are lower using this current analysis and
the measured 90\% C.L. branching fraction upper limits are less
stringent compared to the previous results by between 3\% and 80\%,
depending on the mode. This is compatible with the use here of a
generic selection procedure for the eleven reported modes.

The results of the ML fits to the on-resonance data are summarized in
Table~\ref{tab:summary}. The signal significance is defined as
$\calS=\sqrt{2\Delta\ln {\cal L}}$, where $\Delta\ln {\cal L}$ is the
change in log-likelihood from the maximum value to the value when the
number of signal events is set to zero. Systematic errors are included
in the $\ln {\cal L}$ distribution by convolving the likelihood
function with a Gaussian distribution with a standard deviation equal
to the total systematic uncertainty, defined later in this article. If
the log-likelihood corresponds to a negative signal, we assign a
significance of zero. The branching fraction \calB\ is given by
$n_{\rm s}/(\eta N_{\BB})$, where $n_{\rm s}$ is the signal yield,
corrected for the fit bias, $\eta$ is the reconstruction efficiency,
and $N_{\BB}$ is the number of \BB\ pairs collected. We assume equal
production rates of \BpBm\ and \BzBzb\ mesons.

\begin{table*}[htbp!]
\caption[Summary of results] {Summary of results for the measured \B
  decay modes: total number of events in analysis region, signal yield
  $n_{\rm s}$ (corrected for fit bias) and its
  statistical uncertainty, reconstruction efficiency $\eta$, daughter
  branching fraction product $\Pi \calB_i (\%)$, significance $S$
  (systematic uncertainties included), measured branching fraction
  \calB, and the 90\% C.L. upper limit ($\calB_{UL}$).}
\begin{center}
\resizebox{1.0\textwidth}{!}{
\begin{tabular}{lrrrrcrr}
\hline\hline
\noalign{\vskip1pt}
Mode & Events & \multicolumn{1}{c}{Yield} &
\multicolumn{1}{c}{$\eta$(\%)} & $\Pi\calB_i(\%)$ & S($\sigma$) & 
\multicolumn{1}{c}{\calB ($\times 10^{-7}$)} & 
$\calB_{UL}$ ($\times 10^{-7}$) \\ \hline
\Bp\to\Kstarm\ep\ep   & & &           &     & 1.2 & $1.7\pm1.4\pm0.1$ & 4.0 \\ 
\,\,\,\,\,\,\Kstarm\to$K^-\pi^0$ & \KstarKpeeNtot & \KstarKpeeNsig & \KstarKpeeEff & 33.3 & \KstarKpeeSigfull & \KstarKpeeBF &  \KstarKpeeUL\\ 
\,\,\,\,\,\,\Kstarm\to\KS\pim & \KstarKzeeNtot & \KstarKzeeNsig & \KstarKzeeEff & 22.8 & \KstarKzeeSigfull & \KstarKzeeBF &  \KstarKzeeUL\\ 
\Bp\to\Kstarm\ep\mup  & & &           &      & 0.0 & $-4.5\pm2.6\pm0.4$  & 3.0 \\ 
\,\,\,\,\,\,\Kstarm\to$K^-\pi^0$ & \KstarKpemuNtot & \KstarKpemuNsig & \KstarKpemuEff & 33.3 & \KstarKpemuSigfull & \KstarKpemuBF & \KstarKpemuUL \\ 
\,\,\,\,\,\,\Kstarm\to\KS\pim & \KstarKzemuNtot & \KstarKzemuNsig & \KstarKzemuEff & 22.8 & \KstarKzemuSigfull & \KstarKzemuBF & \KstarKzemuUL \\ 
\Bp\to\Kstarm\mup\mup  & & &           &      & 1.3 & $2.4\pm1.8\pm0.4$   & 5.9 \\ 
\,\,\,\,\,\,\Kstarm\to$K^-\pi^0$ & \KstarKpmumuNtot & \KstarKpmumuNsig & \KstarKpmumuEff & 33.3 & \KstarKpmumuSigfull & \KstarKpmumuBF & \KstarKpmumuUL \\ 
\,\,\,\,\,\,\Kstarm\to\KS\pim & \KstarKzmumuNtot & \KstarKzmumuNsig & \KstarKzmumuEff & 22.8 & \KstarKzmumuSigfull & \KstarKzmumuBF & \KstarKzmumuUL \\ 
\BtoRhoee   & \RhoeeNtot & \RhoeeNsig & \RhoeeEff       & 100.0 & \RhoeeSigfull & \RhoeeBF & \RhoeeUL \\ 
\BtoRhoemu  & \RhoemuNtot & \RhoemuNsig & \RhoemuEff    & 100.0 & \RhoemuSigfull & \RhoemuBF & \RhoemuUL \\ 
\BtoRhomumu & \RhomumuNtot & \RhomumuNsig & \RhomumuEff & 100.0 & \RhomumuSigfull & \RhomumuBF & \RhomumuUL \\ 
\BtoDcee   & \DceeNtot & \DceeNsig & \DceeEff       & 9.13 & \DceeSigfull & \DceeBF & \DceeUL \\ 
\BtoDcemu  & \DcemuNtot & \DcemuNsig & \DcemuEff    & 9.13 & \DcemuSigfull & \DcemuBF & \DcemuUL \\ 
\BtoDcmumu & \DcmumuNtot & \DcmumuNsig & \DcmumuEff & 9.13 & \DcmumuSigfull & \DcmumuBF & \DcmumuUL \\ 
%
%\BtoPee & \PeeNsig & \PeeEff & 100.0 & \PeeSigfull & \PeeBF & \PeeUL \\ 
%\BtoPemu & \PemuNsig & \PemuEff & 100.0 & \PemuSigfull & \PemuBF & \PemuUL \\ 
%\BtoPmumu & \PmumuNsig & \PmumuEff & 100.0 & \PmumuSigfull & \PmumuBF & \PmumuUL \\ 
%
\BtoKemu  & \KemuNtot & \KemuNsig & \KemuEff & 100.0 & \KemuSigfull & \KemuBF & \KemuUL \\ 
\BtoPiemu & \PiemuNtot & \PiemuNsig & \PiemuEff & 100.0 & \PiemuSigfull & \PiemuBF & \PiemuUL \\ 

\hline\hline
\end{tabular}
}
\end{center}
\label{tab:summary}
\end{table*}

\begin{figure}[hbt!] 
\begin{center} 
\begin{tabular}{c}
    \epsfig{file=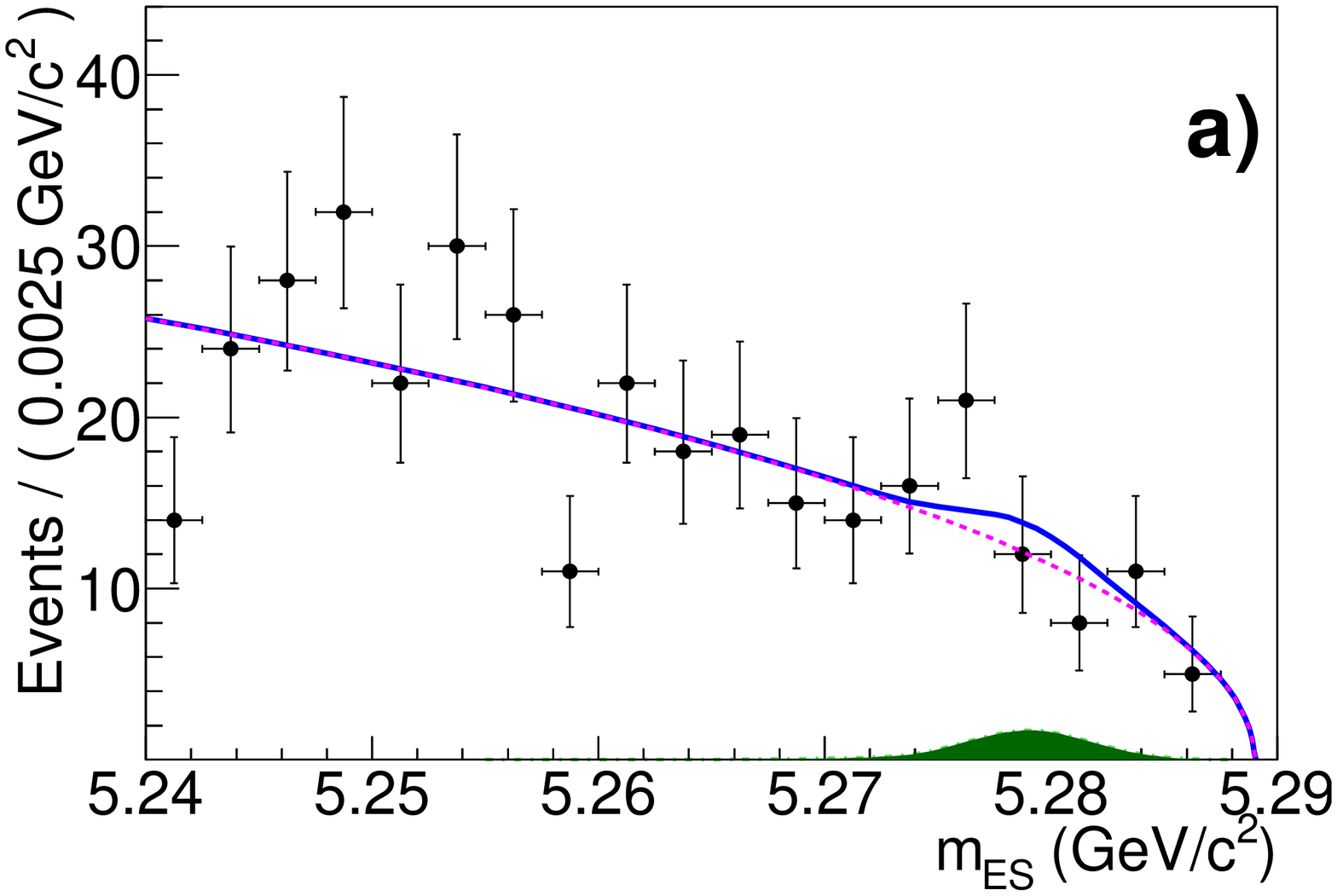,width=0.80\columnwidth}  \\
    \epsfig{file=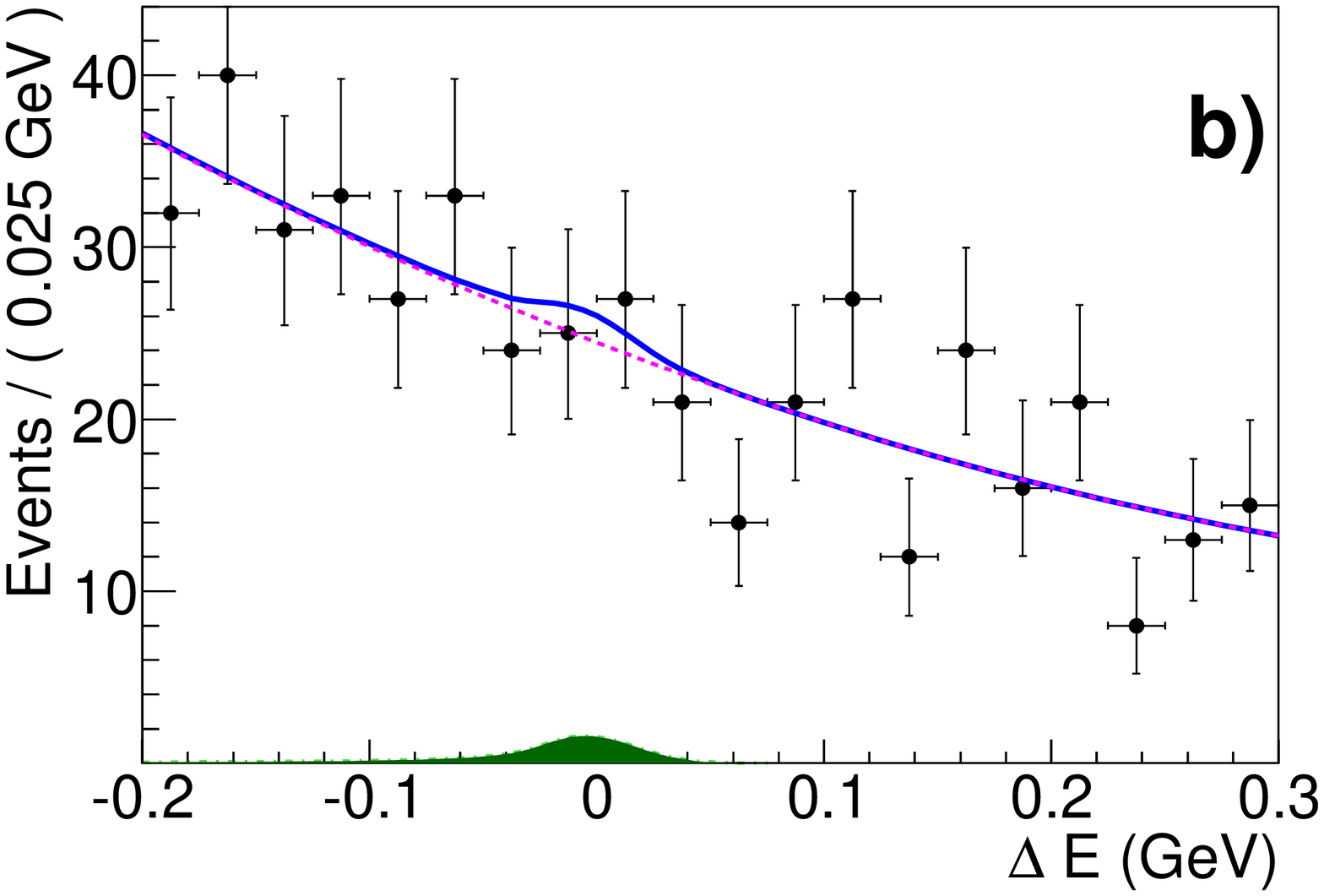,width=0.80\columnwidth} \\
    \epsfig{file=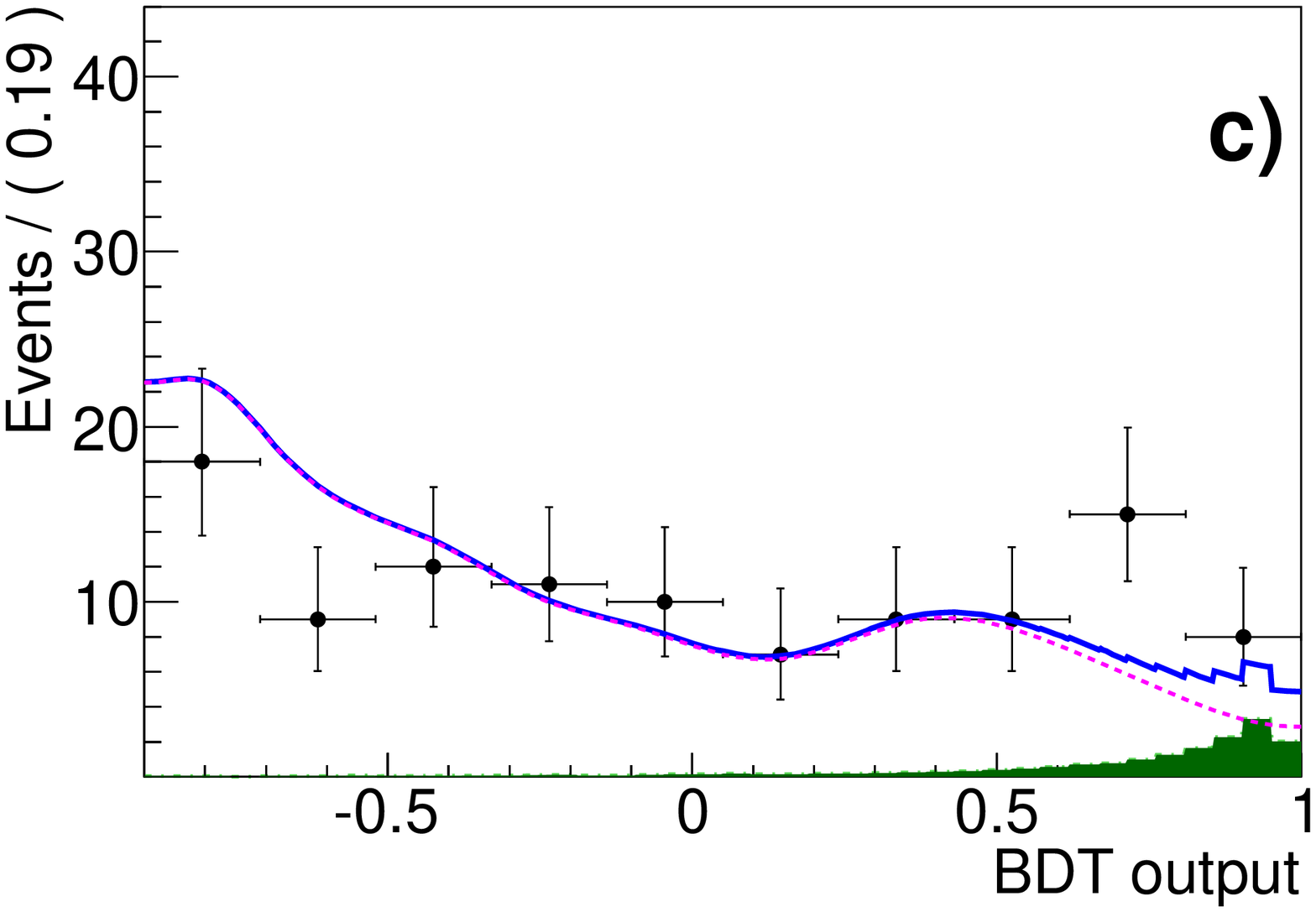,width=0.80\columnwidth}
\end{tabular}
\caption{Projections of the multidimensional fit onto a) \mes; b)
  \DeltaE; and c) \lhratio\ output for the mode $\Bp\to\pim\ep\mup$. The
  points with error bars show the data; the dashed line is the
  background PDF; the solid line is the signal-plus-background PDF;
  and the solid area is the signal PDF.}
\label{fig:fig02}
\end{center} 
\end{figure}

\begin{figure}[hbt!] 
\begin{center} 
\begin{tabular}{c}
    \epsfig{file=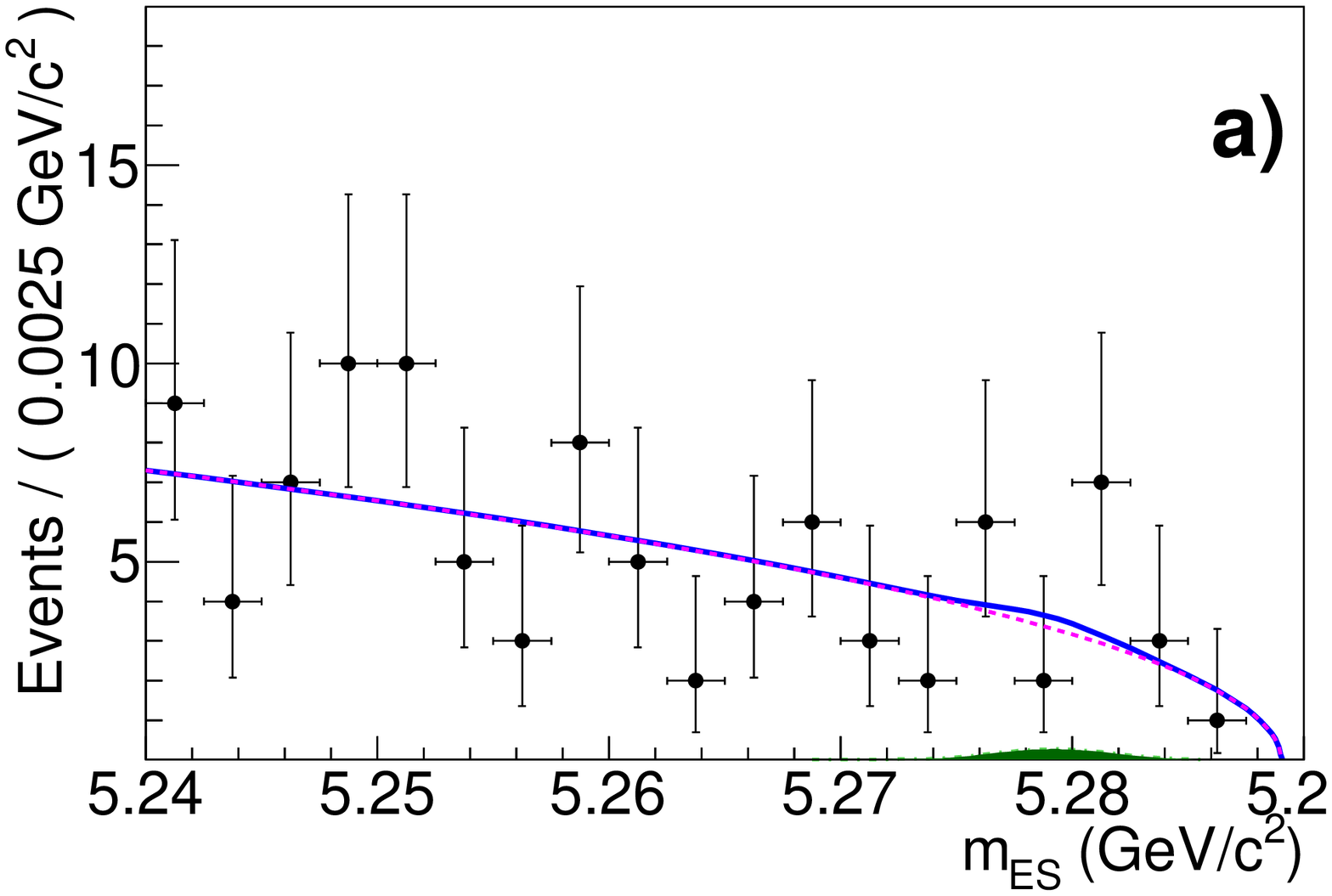,width=0.80\columnwidth} \\
    \epsfig{file=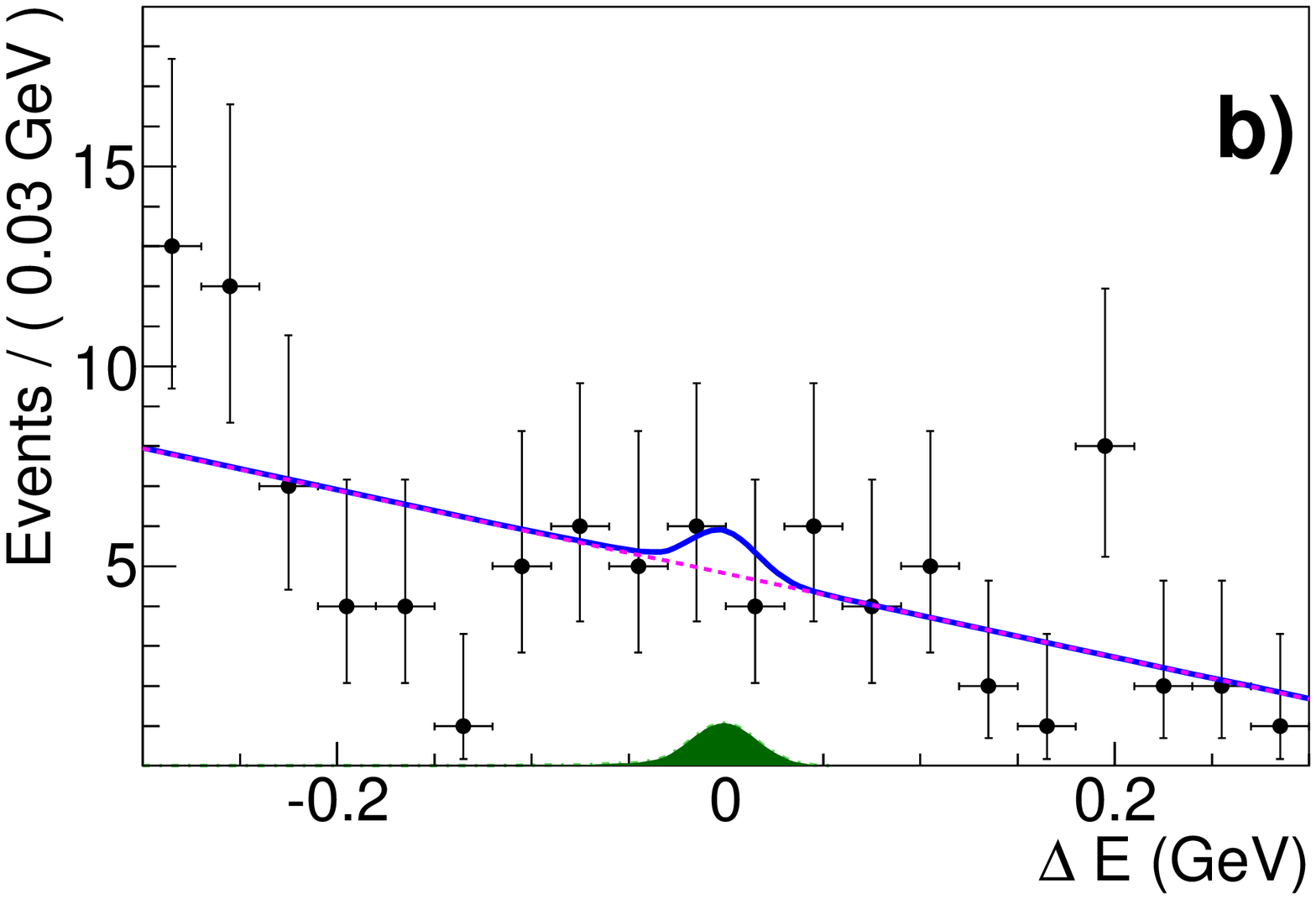,width=0.80\columnwidth} \\
    \epsfig{file=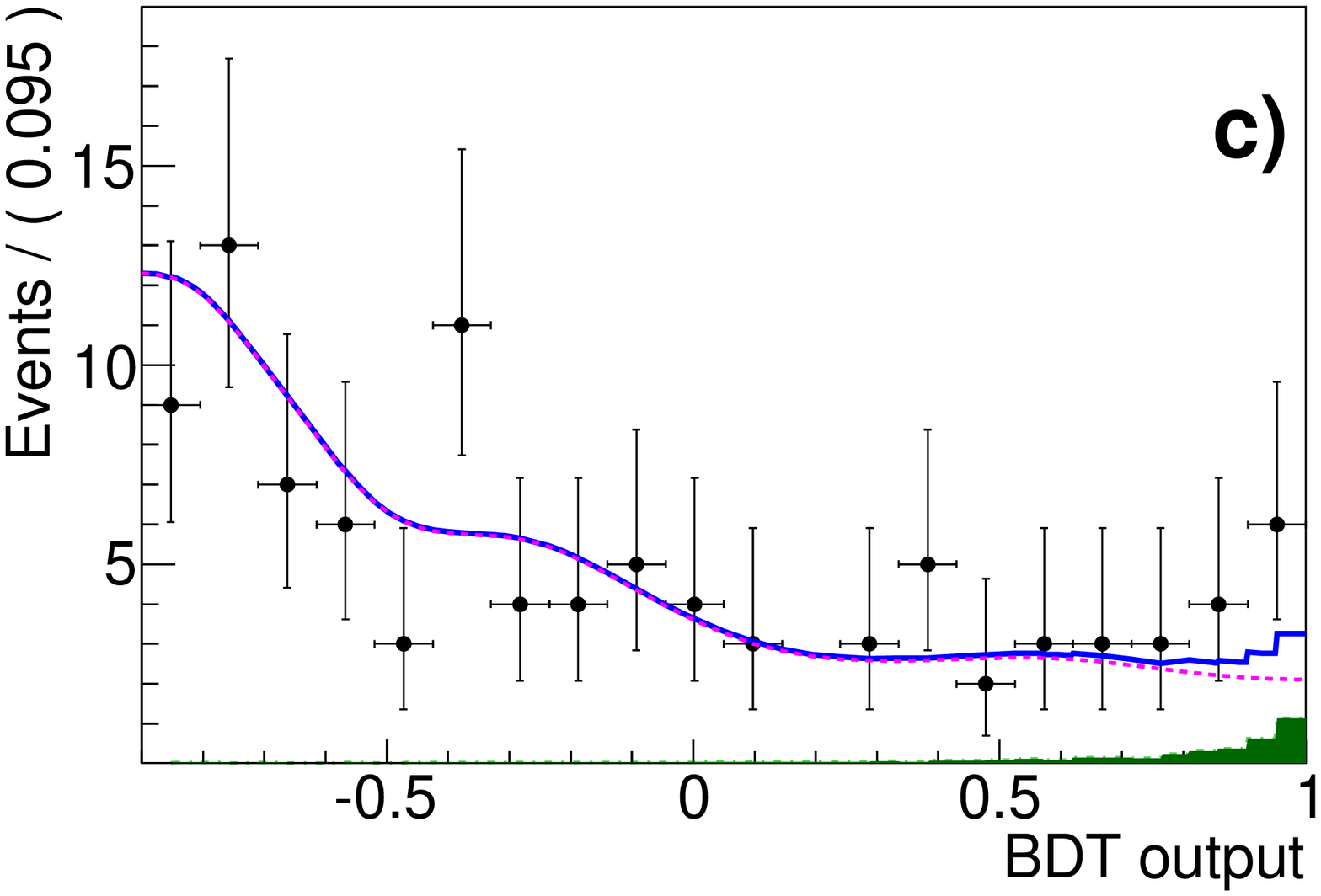,width=0.80\columnwidth} \\
    \epsfig{file=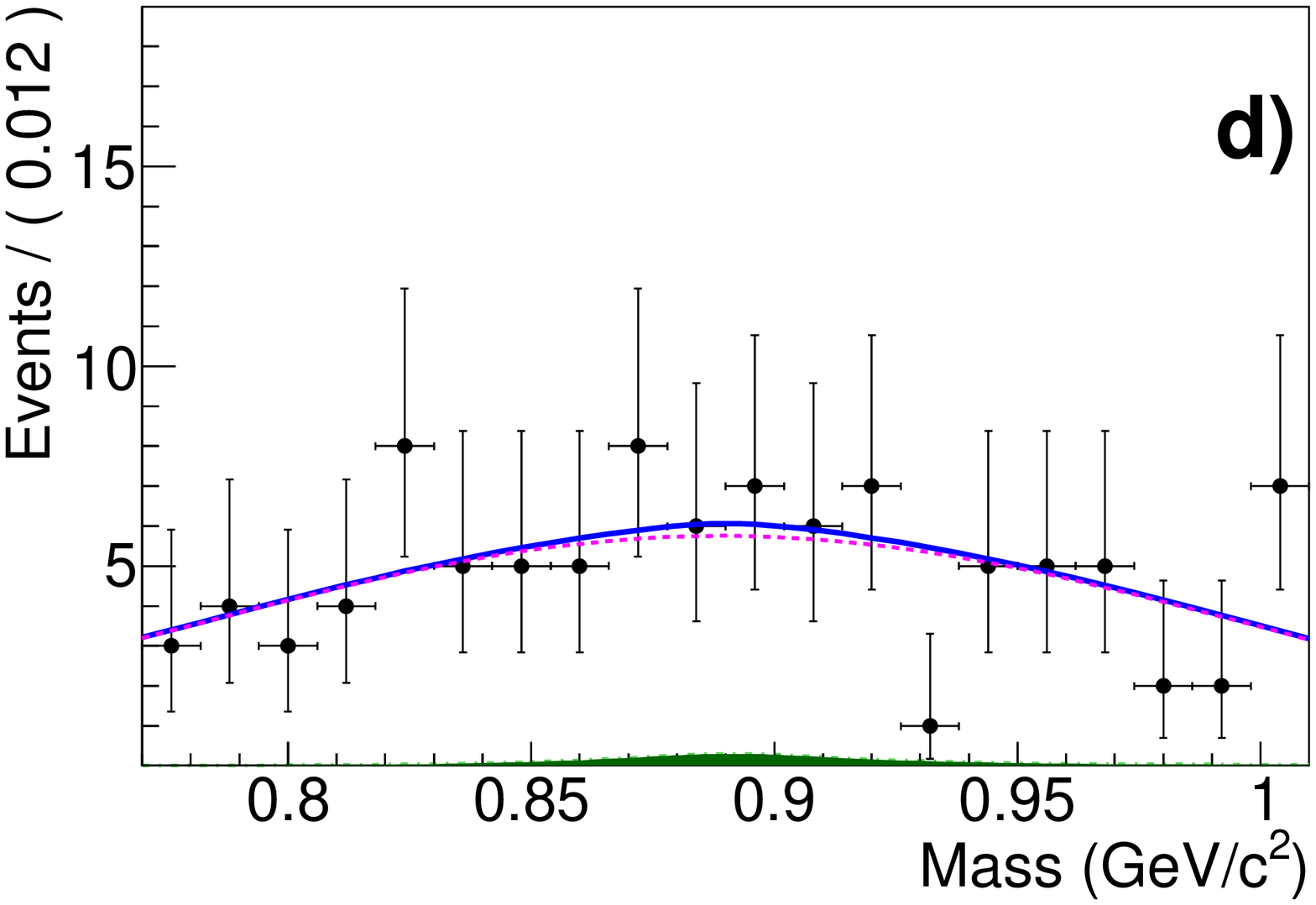,width=0.80\columnwidth}
\end{tabular}
\caption{Projections of the multidimensional fit onto a) \mes; b)
  \DeltaE; c) \lhratio\ output; and d) mass for the mode
  $\Bp\to\Kstarm\mup\mup$ ($\Kstarm\to\KS\pim$). The
  points with error bars show the data; the dashed line is the
  background PDF; the solid line is the signal-plus-background PDF;
  and the solid area is the signal PDF.}
\label{fig:fig03}
\end{center} 
\end{figure}

Figures~\ref{fig:fig02} and~\ref{fig:fig03} show the projections of the fit onto
the discriminating variables for two of the modes,
$\Bp\to\pim\ep\mup$ and $\Bp\to\Kstarm\mup\mup$ ($\Kstarm\to\KS\pim$). 
The candidates in the figure are subject to the
requirement on the probability ratio ${\cal P}_{\rm sig}/({\cal
P}_{\rm bck}+{\cal P}_{\rm sig})>0.9$, where ${\cal P}_{\rm sig}$ and
${\cal P}_{\rm bck}$ are computed without the use of the variable
plotted. The other modes show similar distributions.

The systematic uncertainties in the branching fractions arise from the
PDF parameterization, fit biases, background yields, and efficiencies.
The PDF uncertainties are calculated by varying, within their errors, the
PDF parameters that are held fixed in the default fit, taking into
account correlations.  For the KEYS algorithm, we vary the smearing
parameter between 50\% and 150\% of the nominal value~\cite{bib:keys},
and for the histograms we change the number of bins used. 
The uncertainty for the fit bias includes the statistical uncertainty
in the mean difference between the fitted signal yield from the ensemble
of 10,000 MC datasets described above and the signal yield from the
fit to the default MC sample, and half of the correction itself, added
in quadrature.

To calculate the contribution to the uncertainty caused by the
assumption that the \qqbar, \ccbar, and \BB\ backgrounds have similar
distributions, we first vary the relative proportions of \qqbar,
\ccbar, and \BB\ used in the simulated background between 0\% and
100\% and retrain the \lhratio\ function for each variation. The new
simulated background BDT PDF is then used in the fit to the data and
the fitted yields compared to the default fit to data. The uncertainty
is taken to be half the difference between the default fit and the
maximum deviation seen in the ensemble of fits. All the uncertainties
described previously are additive in nature and affect the
significance of the branching fraction results. The total additive
signal yield uncertainty is between 0.2 and 0.7 events, depending on
the mode.

The sources of multiplicative uncertainties include: reconstruction
efficiency from tracking (0.8\% per track for the leptons and 0.7\%
for the kaon or pion, added linearly); neutral \piz\ and \KS\
reconstruction efficiency (3.0\% and 1.0\%, respectively); charged
particle identification (0.7\% for electrons, 1.0\% for muons, 0.2\%
for pions, 1.1\% for kaons, added linearly); the BDT
response from comparison to charmonium control samples such as
$\Bm\to\jpsi X^-$ (2.0\%); and the number of \BB\ pairs
(0.6\%)~\cite{BaBarDetector}. The total multiplicative branching
fraction uncertainty is 5\% or less for all modes.

When forming the overall branching fraction for the
$\Bp\to\Kstarm\ellp\ellpp$ decays, we assume that the overall \Kstarm\
sub-mode additive uncertainties are uncorrelated and the
multiplicative uncertainties are correlated.

As shown in Table~\ref{tab:summary}, we observe no significant yields.
We use a Bayesian approach to calculate the 90\% C.L. branching
fraction upper limits $\calB_{UL}$ by multiplying the likelihood
distributions with a prior which is null in the unphysical regions
($n_{\rm s}<0$) and constant elsewhere. The total likelihood distribution
is integrated (taking into account statistical and systematic
uncertainties) as a function of the branching fraction from 0 to
$\calB_{UL}$, such that $\int^{\calB_{UL}}_{0} \calL\, d\calB =
0.9\times\int^{\infty}_{0} \calL\, d\calB$. For the overall 
\Kstarm\ellp\ellpp\ results, the total likelihood distributions for
the two sub-modes are first combined before integration. The
upper limits in all cases are dominated by the statistical
uncertainty.

In summary, we have searched for eleven lepton-number violating
processes $\Bp\to X^-\ellp\ellpp$. We find no significant yields and
place 90\% C.L. upper limits on the branching fractions in the range
$(1.5-26)\times 10^{-7}$. The limits for the modes with a \rhom,
\pim or \Km\ are an order of magnitude more stringent than previous
best measurements~\cite{bib:cleo}. The limits for the
\Bp\to\Dm\ellp\ellpp\ are compatible with those reported in
Ref.~\cite{bib:lnv_belle}.

We are grateful for the excellent luminosity and machine conditions
provided by our \pep2\ colleagues, 
and for the substantial dedicated effort from
the computing organizations that support \babar.
The collaborating institutions wish to thank 
SLAC for its support and kind hospitality. 
This work is supported by
DOE
and NSF (USA),
NSERC (Canada),
CEA and
CNRS-IN2P3
(France),
BMBF and DFG
(Germany),
INFN (Italy),
FOM (The Netherlands),
NFR (Norway),
MES (Russia),
MICIIN (Spain),
and STFC (United Kingdom). 
Individuals have received support from the
Marie Curie EIF (European Union)
and the A.~P.~Sloan Foundation (USA).

%------------------------------------------------------------


\begin{thebibliography}{99}
\bibitem{bib:charge}
Charge-conjugate modes are implied throughout this paper.
\bibitem{bib:cruijff}
The Cruijff function is a centered Gaussian with different left-right
resolutions and non-Gaussian tails: $f(x) = \exp((x-m)^2/(2\sigma^2_{L,R} + \alpha_{L,R}(x-m)^2)).$
\end{thebibliography}
\end{document}